\documentclass[
floatfix,
aps,
pra,
amsmath,
twocolumn,
superscriptaddress
]{revtex4-1}

\pdfoutput=1

\usepackage{graphicx}
\usepackage{wrapfig}
\usepackage{amsmath}
\usepackage{amssymb}
\usepackage{bbm}
\usepackage{hyperref}

\begin{document}

\title{Microwave-activated controlled-$Z$ gate for fixed-frequency fluxonium qubits}

\author{Konstantin N. Nesterov}
\affiliation{Department of Physics, University of Wisconsin-Madison, Madison, WI 53706}
\author{Ivan V. Pechenezhskiy}
\affiliation{Department of Physics, Joint Quantum Institute,
and 
Center for Nanophysics and
\\ Advanced Materials,
University of Maryland, College Park, MD 20742}
\author{Chen Wang}
\affiliation{Department of Physics, University of Massachusetts, Amherst, MA 01003}
\author{Vladimir E. Manucharyan}
\affiliation{Department of Physics, Joint Quantum Institute,
and 
Center for Nanophysics and
\\ Advanced Materials,
University of Maryland, College Park, MD 20742}

\author{Maxim G. Vavilov}
\affiliation{Department of Physics, University of Wisconsin-Madison, Madison, WI 53706}

\date{\today}

\begin{abstract}
The superconducting fluxonium circuit is an artificial atom with a strongly anharmonic spectrum: When biased at a half flux quantum, the  lowest qubit transition is an order of magnitude smaller in frequency than those to higher levels. Similar to conventional atomic systems, such a frequency separation between the computational and noncomputational subspaces allows independent optimizations of the qubit coherence and two-qubit interactions. Here, we describe a controlled-$Z$ gate for two fluxoniums connected either capacitively or inductively, with qubit transitions fixed near $500~\textrm{MHz}$. The gate is activated by a microwave drive at a resonance involving the second excited state. We estimate intrinsic gate fidelities over 99.9\% with gate times below 100 ns.
\end{abstract}

\pacs{}

\maketitle

The transmon qubit~\cite{Koch2007_pra} has been largely responsible for the recent spectacular breakthrough in superconducting quantum information processing~\cite{Devoret2013_science}. This qubit shows long coherence times~\cite{Rigetti2012_prb, Sheldon2016_pra}, high-fidelity gates~\cite{Chow2009_prl}, and reliable  readout techniques~\cite{Houck2008_prl, Campagne-Ibarcq2013_prx, Hatridge2013_science}.   Although multiqubit devices~\cite{Takita2016_prl, Neill2018_science, Song2017_prl} with efficient two-qubit gates~\cite{DiCarlo2009_nature, McKay2016_prappl, Sheldon2016_pra, Paik2016} have already been demonstrated, the weak anharmonicity of the transmon presents a substantial challenge in pushing the fidelities higher. Fundamentally, the main issue is that both the qubit memory and interaction with other qubits are done using transitions with nearly identical frequencies and matrix elements. This prevents decoupling of a transmon from its dissipative environment without  increasing the gate time. Furthermore, stronger coupling of two transmons requires a smaller detuning of their frequencies, which in turn enhances the  state leakage outside of the computational subspace.

In atomic systems, qubit states are chosen in such a way that the transition between them is forbidden by the selection rules to provide long coherence in the computational subspace.  Quantum  gates and qubit readout are performed through transitions outside of  that subspace with a stronger coupling to electromagnetic fields.  Such a separation of  states for information storage and processing allows one to perform many high-fidelity gates before the qubit state is spoiled by decoherence. This was realized in architectures based on nitrogen-vacancy centers~\cite{Dutt2007_science}, trapped ions~\cite{Vittorini2014_pra}, and Rydberg atoms~\cite{Maller2015_pra}. In superconducting systems, a similar idea was implemented in experiments where the qubit quantum state is stored in a high-quality microwave resonator~\cite{Mariantoni2011_science, Rosenblum2018_natcomm}, while the physical superconducting qubits are used only for short times during gate realizations.

In this paper, we consider the fluxonium superconducting qubit~\cite{Manucharyan2009_science, Manucharyan2012_prb, Lin2018_prl} biased at a half of magnetic-flux quantum, which can be utilized as both a long memory and a fast processor of quantum information. 
The fluxonium is topologically distinct from the transmon and thereby combines strong Josephson nonlinearity with complete insensitivity to offset charges~\cite{Manucharyan2009_science, Koch2009_prl}. As a result, the fluxonium shares many spectral features with a multi-level atomic system. 
Its two lowest energy states can have a very long coherence time~\cite{Manucharyan2012_prb, Manucharyan2018} and thus are well suited for quantum information storage.  
Higher energy states are separated by much larger energy spacings  and have large transition matrix elements similar to a typical transmon, which makes them ideal for the information processing~\cite{Manucharyan2012_prb, Zhu2013_prb2, Lin2018_prl}. We propose a controlled-$Z$ (CZ) gate for two fluxoniums with an always-on interaction realized through  either electrostatic or inductive coupling.
The gate is activated by a microwave pulse, while the qubits are kept at  fixed frequencies. 
By numerically simulating the driven dynamics  and taking into account  excited states,  we
 find that the intrinsic gate fidelity can be above 99.9\% for  gate times below 100 ns for realistic coupling parameters when decoherence processes are neglected.  Below, we will discuss the effect of decoherence on the gate fidelity.

\begin{figure}[t]
\includegraphics[width=0.95\columnwidth]{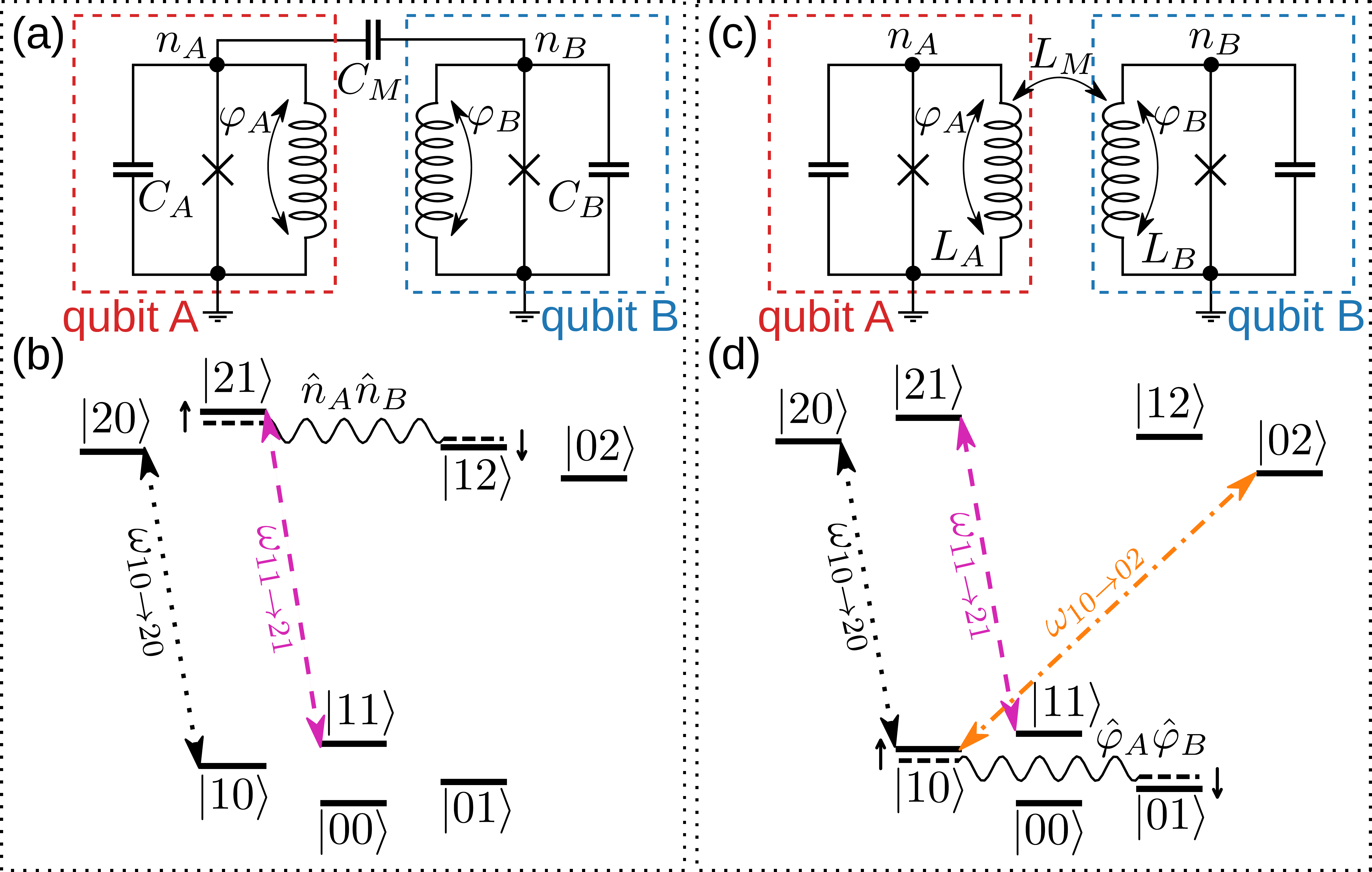}\caption{(Color online) 
Left panel: (a) The circuit diagram of two fluxoniums coupled by a capacitor  $C_M$. (b)   Energy levels (horizontal solid lines) of the interacting system. The primary effect of coupling is to induce the level repulsion (marked by vertical arrows) between noninteracting states $|12\rangle_0$ and $|21\rangle_0$  (horizontal dashed lines). The repulsion results in  $\omega_{10\to 20}\neq \omega_{11\to 21}$, and the gate is realized by a resonant drive of the $|11\rangle \to |21\rangle$ transition. Right panel (the notation follows the left panel): (c) Two fluxoniums coupled by a mutual inductance $L_M$. (d) The flux-flux interaction $\hat\varphi_A\hat\varphi_B$ primarily affects states $|10\rangle$ and $|01\rangle$, leads to $\omega_{10\to 20}\neq \omega_{11\to 21}$, and permits the $|10\rangle \to |02\rangle$ transition  (dashed-dotted line).
}\label{Fig-fluxonium}
\end{figure}

\textit{Fluxonium at half-flux-quantum bias.}  
Each fluxonium circuit [dashed boxes in Figs.~\ref{Fig-fluxonium}(a) and \ref{Fig-fluxonium}(c)] features a long Josephson-junction array with the total inductance $L_\alpha$ shunting a phase-slip Josephson junction,
where $\alpha=A$ or $B$ is a  qubit label in a two-qubit device.
The circuit is characterized by three energy scales: the charging energy $E_{C, \alpha} = e^2/2C_\alpha$, the inductive energy $E_{L,\alpha} = (\hbar/2e)^2/L_\alpha$, and the Josephson energy $E_{J, \alpha}$,  where $-e$ is the electron charge, $C_\alpha$ is the total capacitance,  and $\hbar = h/2\pi$ is the Planck constant. With an external flux $\Phi_{\rm ext, \alpha} = (\hbar/2e)\phi_{\rm ext, \alpha}$ threading the loop formed by the Josephson junction and the inductance, the Hamiltonian of fluxonium $\alpha$ is~\cite{Manucharyan2009_science}
\begin{equation}\label{Hamiltonian-fluxonium}
 \hat{H}_{\alpha}^{(0)} = 4E_{C,\alpha} \hat{n}_\alpha^2 + \frac 12 E_{L,\alpha} \hat{\varphi}_\alpha^2 - E_{J,\alpha} \cos(\hat{\varphi}_\alpha - \phi_{\rm ext,\alpha})\,.
\end{equation}
Here, $\hat{\varphi}_\alpha$ and $\hat{n}_\alpha$ are the generalized flux and charge (Cooper-pair number) operators satisfying $[\hat{\varphi}_\alpha, \hat{n}_\alpha] = i$.
There is a wide  experimentally attainable parameter space suitable for the proposed CZ gate. For illustration purposes,
 we use the following  parameters: $E_{C, A}/h = 1.5$ GHz, $E_{C, B}/h = 1.2$ GHz,  $E_{J, A}/h = 5.5$ GHz,  $E_{J, B}/h = 5.7$ GHz, and $E_{L, A}/h= E_{L, B}/h = 1$ GHz.  Unlike in transmons, $E_{J,\alpha}/E_{C,\alpha}$ is not required to be large because the fluxonium is insensitive to the charge noise~\cite{Koch2009_prl}. 

At $\varphi_{\rm ext, \alpha} = \pi$, the fluxoniums are at their sweet spots with respect to the flux noise~\cite{Manucharyan2012_prb}.
In this case,  the splitting between energies $\varepsilon_0^\alpha$ and $\varepsilon_1^\alpha$ of the two lowest levels $|0_\alpha\rangle$ and $|1_\alpha\rangle$ is determined by tunneling between two potential wells (see Fig.~10 in Ref.~\cite{Manucharyan2012_prb}). This gives a relatively small $\omega^\alpha_{0\rightarrow 1}/2\pi \sim 500$ MHz, where $\omega^\alpha_{i\to f} = (\varepsilon_f^\alpha - \varepsilon_i^\alpha)/\hbar$ is the frequency of the $|i_\alpha\rangle\rightarrow |f_\alpha\rangle$ transition.  On the other hand, a higher frequency $\omega^\alpha_{1\to 2}/2\pi \sim 5$ GHz facilitates strong coupling and fast gates using the second excited state $|2_\alpha\rangle$.

\textit{Capacitive coupling.}
First, we consider  coupling realized via a capacitor $C_M$ as shown in Fig.~\ref{Fig-fluxonium}(a). 
The Hamiltonian of the two-qubit system is
\begin{equation}\label{Hamiltonian-two-qubit}
\hat{H} = \hat{H}^{(0)}_{A} + \hat{H}^{(0)}_B + \hat{V} + \hat{H}_{\rm drive}\,,
\end{equation}
where $\hat{V}$ is the qubit-qubit interaction and $\hat{H}_{\rm drive}$ describes the  external microwave drive. 
In the limit $C_M \ll C_A, C_B$, we have~\cite{Vool2017_circuit_theory}
\begin{equation}\label{interaction-charge}
 \hat{V} = J_C \hat{n}_A \hat{n}_B \quad{\rm with}\quad J_C = 4e^2{C_M}/\left({C_A C_B}\right)\,,
\end{equation}
where $\hat{n}_{\alpha}$ is the total charge on $C_\alpha$ and the corresponding side of $C_M$.
We note that $C_M$ also slightly renormalizes $E_{C,\alpha}$.
For simplicity, we assume that the drive is directly applied to the qubits with effective couplings $\eta_A$ and $\eta_B$,
\begin{equation}\label{drive}
 \hat{H}_{\rm drive} =f(t)\cos(\omega_d t) \left(\eta_A \hat{n}_A + \eta_B \hat{n}_B\right) \,,
\end{equation}
where $f(t)$ describes a pulse shape.
We use the notation  $|kl\rangle$ for an eigenstate of $\hat{H}$ with $\hat{H}_{\rm drive}=0$, which is adiabatically connected to the noninteracting eigenstate $|kl\rangle_0 = |k_A\rangle |l_B\rangle$. We denote the frequencies of two-qubit transitions as $\omega_{kl\to k'l'}$.

The main idea of the CZ gate is as follows.
When $\hat{V}=0$,  $\omega_{10\rightarrow 20}= \omega_{11\rightarrow 21} = \omega^A_{1\to 2}$. 
The interaction $\hat{V}$ lifts this degeneracy, and the gate can be realized by selectively driving Rabi oscillations between states $|11\rangle$ and $|21\rangle$, see Fig.~\ref{Fig-fluxonium}(b). State $|11\rangle$ accumulates an extra phase factor of $e^{i\pi}$ after the system completes one oscillation. 
Therefore, for the ideal pulse shape $f(t)$, when other transitions, in particular, $|10\rangle \rightarrow |20\rangle$, are not  affected, 
the computational subspace $\{|00\rangle, |01\rangle, |10\rangle, |11\rangle\}$ would evolve according to the CZ gate operator 
 $\hat{U}_{\rm CZ} = {\rm diag}(1, 1, 1, -1)$
up to single-qubit $Z$ gates~\cite{Ghosh2013_pra}.

 At $\varphi_{\rm ext, \alpha} = \pi$, each single-fluxonium level has a well-defined even or odd parity with respect to $\varphi \rightarrow -\varphi$, implying zero matrix elements of $\hat{\varphi}$ and $\hat{n}$ for pairs of levels with the same parity~\cite{Zhu2013_prb1, Zhu2013_prb2}. Thus, $\varphi^\alpha_{0\to 2} = n^\alpha_{0\to 2} = 0$, where  $O^\alpha_{i\to f} = |\langle i_\alpha |\hat{O}_\alpha|f_\alpha \rangle|$ is the magnitude of the single-qubit matrix element ($\hat{O} = \hat{\varphi}$ or $\hat{n}$). 
 Furthermore, $ n^\alpha_{0\to 1} \ll n^\alpha_{1\to 2}$ since $n^\alpha_{0\to 1}$ is suppressed by weak tunneling between two potential wells.  
 This matrix-elements hierarchy has two important consequences: (i) 
$\hat{H}_{\rm drive}$ couples more strongly to the $|1\rangle \rightarrow |2\rangle$ transition than to the $|0\rangle \rightarrow |1\rangle$; (ii) $\hat{V}$ leads to a much stronger hybridization (level repulsion) between noncomputational states $|21\rangle$ and $|12\rangle$ [Fig.~\ref{Fig-fluxonium}(b)] than computational ones because $n^A_{1\to 2} n^B_{1\to 2} \gg n^A_{0\to 1} n^B_{0\to 1}$. 
As a result, the  second-order  correction to the energy of state $|21\rangle$ is by a factor of several hundreds larger than those to the energies  of states $|10\rangle$ and $|11\rangle$. Since $n^\alpha_{0\to 2}  = n^\alpha_{2\to 2} =0$,  the state $|20\rangle_0$ does not acquire the perturbative shift from nearby levels $|02\rangle_0$, $|21\rangle_0$, and $|12\rangle_0$, making $\omega_{11\rightarrow 21}\neq \omega_{10\rightarrow 20}$.

\begin{figure}[t]
\includegraphics[width=0.9\columnwidth]{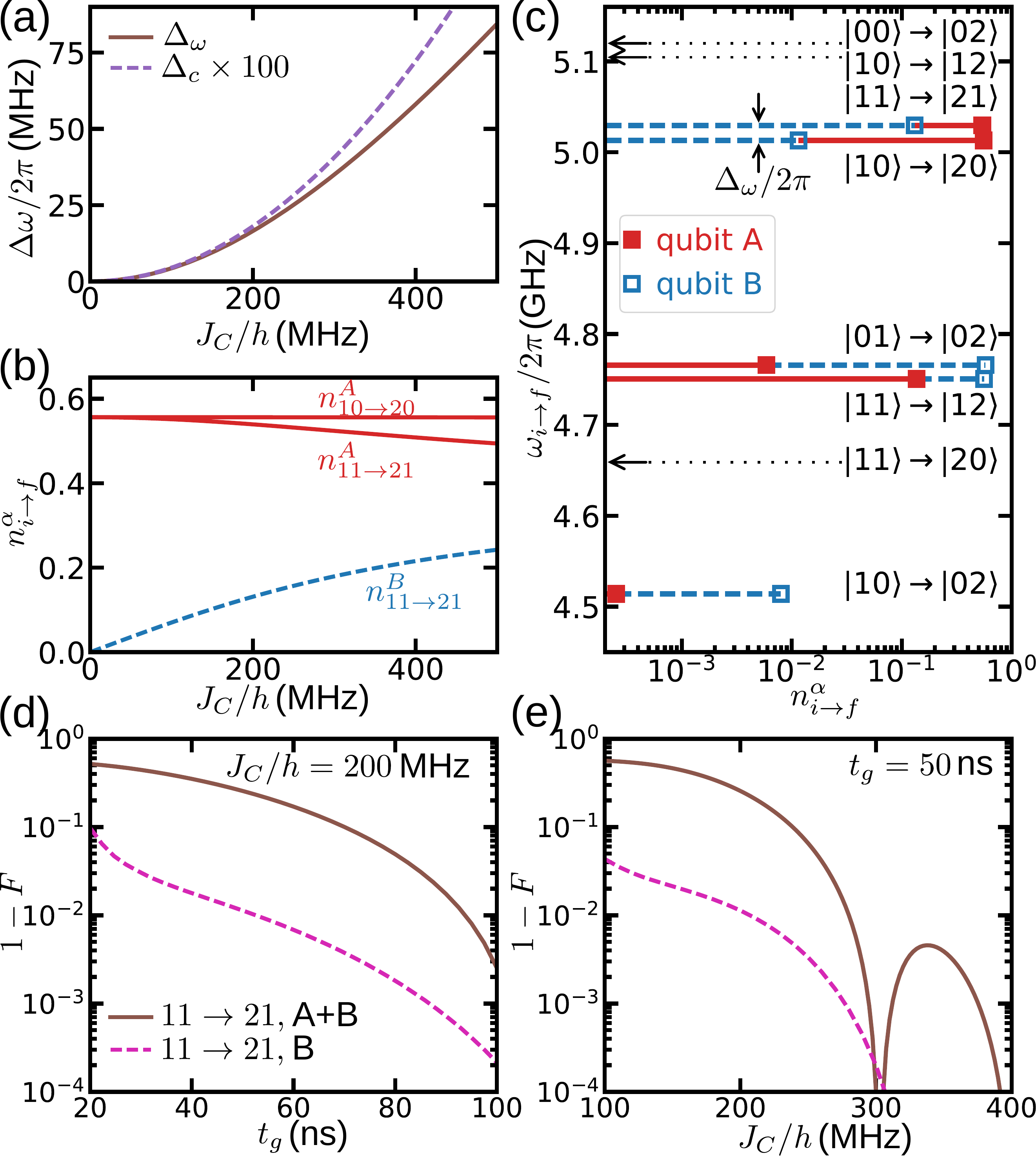}
\caption{(Color online) Numerical results for the capacitive coupling. 
(a) The frequency mismatch $\Delta_\omega$ and the crosstalk parameter $\Delta_c$ vs the interaction strength $J_C$. (b) Amplitudes of the  matrix elements of $\hat{n}_A$ and $\hat{n}_B$ vs $J_C$. (c) Transition frequencies ($y$ axis) and matrix elements ($x$ axis) of $\hat{n}_A$ (solid squares) and $\hat{n}_B$ (open squares)  at $J_C/h = 200$ MHz. The transitions without symbols are prohibited by the symmetry.
(d), (e) The gate error vs the gate time $t_g$ at a fixed $J_C/h = 200$ MHz (d) and vs $J_C$ at $t_g = 50$ ns (e) for the microwave drive applied at $\omega_{11 \rightarrow 21}$ to both qubits (solid lines) and selectively to qubit $B$ only (dashed lines).}\label{Fig-capacitive}
\end{figure}

The gate rate~\cite{Chow2013_njphys}, characterized by the frequency mismatch  $\Delta_\omega = \omega_{11\rightarrow 21}- \omega_{10\rightarrow 20}$, increases with decreasing $\delta =|\omega^A_{1\rightarrow 2} - \omega^B_{1\rightarrow 2}|$, which is $248$ MHz for our choice of parameters. In general, smaller $\delta$ leads to faster and better performing gates. Our conservative choice of $\delta \sim 250$~MHz allows room for a frequency selectivity of multiple neighbors in a larger-scale system.

Taking five levels in each qubit, we numerically diagonalize the two-qubit Hamiltonian (\ref{Hamiltonian-two-qubit}) at $\hat{H}_{\rm drive}=0$  and study $\Delta_\omega$ and $n^\alpha_{kl\to k'l'} = |\langle kl |\hat{n}_\alpha|k'l'\rangle|$ as a function of $J_C$, see Figs.~\ref{Fig-capacitive}(a) and (b). As expected, $\Delta_\omega$ increases monotonically with increasing $J_C$ [solid line in Fig.~\ref{Fig-capacitive}(a)]. We notice that $\Delta_\omega$ increases much faster than  $\Delta_c = \omega_{00 \rightarrow 01} - \omega_{10\rightarrow 11}$, which characterizes the crosstalk in the computational subspace. For our parameters, $\Delta_\omega/\Delta_c\approx 100$.
The matrix elements panel [Fig.~\ref{Fig-capacitive}(b)] illustrates strong mixing between $|12\rangle_0$ and $|21\rangle_0$. While $n^A_{10 \to 20} \approx n^A_{1\to 2}$ at finite $J_C$, the value of $n^A_{11\to 21}$ decreases with increasing $J_C$ since the dressed state $|21\rangle$ is a superposition of noninteracting states $|12\rangle_0$ and $|21\rangle_0$. For the same reason, $n^B_{11\to 21}\neq 0$ for $J_C \ne 0$. 

The finite width of the drive spectrum can potentially result in other transitions out of the computational subspace with frequencies close to $\omega_{11\to 21}$  [Fig.~\ref{Fig-capacitive}(c)].
 Since $n^B_{11\to 21} \gg n^B_{10\to 20} $, the
undesirable activation of  $|10\rangle \to |20\rangle$ can be suppressed by applying the drive selectively to qubit $B$ ($\eta_A=0$), which is similar to the cross-resonance scheme~\cite{Sheldon2016_pra, Chow2011_prl}. We notice that certain matrix elements $n^\alpha_{kl\to k'l'}$ remain equally zero for $\hat{V} \ne 0$. All two-qubit levels $|kl\rangle$ can be separated into two families depending on the parity of $k+l$. To higher orders, $\hat{V}$ mixes only levels with the same parity of $k+l$, while $\hat{n}_\alpha$ only connects levels with different parities of $k+l$. We have $n^\alpha_{10\to 12} =0$ when $J_C \ne 0$, while $n^B_{10\to 02} \ne 0$.

To model the gate operation, we find the unitary evolution operator $\hat{U}(t)$ by integrating numerically $ i\hbar {\partial \hat{U}(t)}/{\partial t} = \hat{H}\hat{U}(t)$. For a desired gate time $t_g$, we consider a Gaussian envelope
$ f(t) = A\{\exp[-8{t(t-t_g)}/{t_g^2}]-1\}$,
where  we later optimize over $A$ and the drive frequency $\omega_d$ within a 15-MHz window around $\omega_{11\to 21}$. 
The evolution operator in the two-qubit computational subspace is represented by a non-unitary $4\times 4$ matrix 
$\hat{U}_{c}$ defined by 
$[\hat{U}_{c}]_{kl,k'l'} = \langle kl|\hat{U}(t_g)|k'l'\rangle$, where $|kl\rangle, |k'l'\rangle \in \{|00\rangle, |01\rangle, |10\rangle, |11\rangle\}$.
To compare $\hat{U}_{\rm c}$ with the ideal operator $\hat{U}_{\rm CZ}$, we apply two instant single-qubit $Z$ rotations to get $\hat{U}_{c}' = \hat{U}_Z \hat{U}_{c}$, where $\hat{U}_Z ={\rm diag}[1, e^{i\delta \phi_{01}}, e^{i\delta\phi_{10}}, e^{i\delta\phi_{01} + i\delta\phi_{10}} ]$, $\delta\phi_{kl} = \phi_{kl} - \phi_{00}$, and $\phi_{kl} =  -{\rm arg} \{[\hat{U}_{c}]_{kl,kl}\}$.
We then calculate the averaged gate fidelity
$F = [{{\rm Tr}(\hat{U}'^\dagger_{c} \hat{U}_{c}') + |{\rm Tr}(\hat{U}^\dagger_{\rm CZ} \hat{U}_{c}')|^2}]/20$~\cite{Pedersen2007_pla, Ghosh2013_pra}.

The gate error $1-F$  is shown in Figs.~\ref{Fig-capacitive}(d) and \ref{Fig-capacitive}(e) as a function of $t_g$ and $J_C$ for the microwave drive applied  to both qubits ($\eta_A=\eta_B = 1$, solid lines, as a worst-case scenario without local microwave control lines for each qubit), and selectively to qubit $B$ ($\eta_A =0, \eta_B =1$, dashed lines). The gate error is higher  for $\eta_A = \eta_B =1$, in which case it is dominated by the $|10\rangle \to |20\rangle$ transition. At sharp minima in solid  lines,  the effect of this transition is negligible. When $\eta_A = 0$, the matrix element $\langle 10 |\hat{H}_{\rm drive} |20\rangle$ is suppressed, providing a higher gate fidelity.
 When the drive is applied to both qubits, $J_C/h$ has to be $\gtrsim 200$ MHz to achieve the 99\% fidelity threshold within $t_g = 100$ ns. For the selective drive of a single qubit, the same interaction strength of 200 MHz leads to the 99\% fidelity for a shorter gate time of 50 ns, and the 99.9\% fidelity is possible at $t_g\approx 90$ ns. 
 The 99.9\% threshold can be achieved at $t_g=50$ ns for both designs provided $J_C$ is sufficiently large. 
 We expect that the gate error can be reduced further with more advanced microwave-pulse shaping~\cite{Motzoi2009_prl}.

\begin{figure}
 \includegraphics[width=0.9\columnwidth]{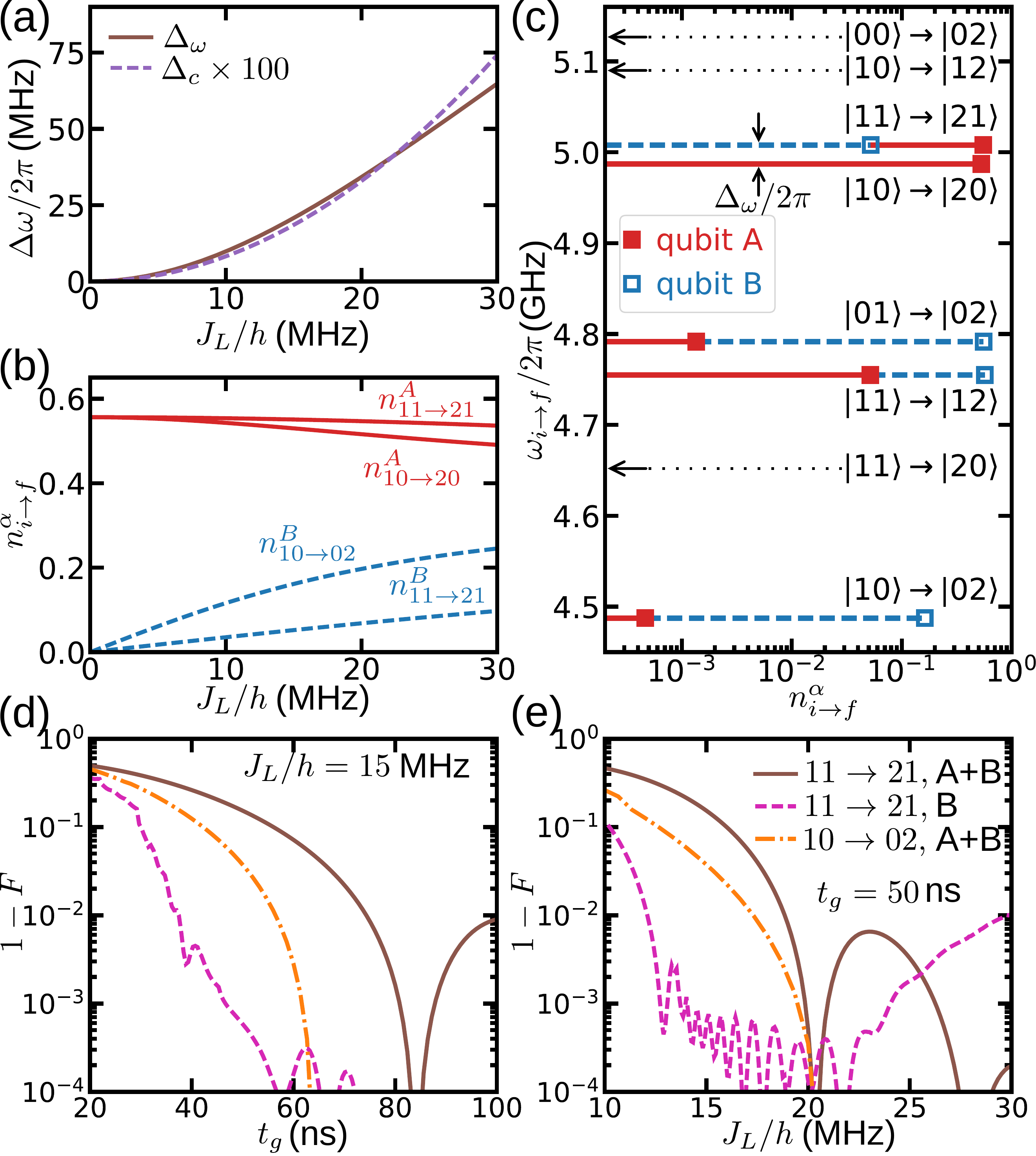}
 \caption{(Color online) Numerical results for the inductive coupling. The notation follows Fig.~\ref{Fig-capacitive} with $J_C$ replaced by $J_L$. The results of panels (c) and (d) are for $J_L/h = 15$ MHz. In addition, in panels (d) and (e), dashed-dotted lines show the results for the drive applied to both qubits at $\omega_{10\rightarrow 02}$. }\label{Fig-inductive}
\end{figure}

\textit{Inductive coupling.}
For the coupling realized via a mutual inductance $L_M \ll L_A, L_B$ as shown in Fig.~\ref{Fig-fluxonium}(c), the interaction $\hat{V}$ in Eq.~(\ref{Hamiltonian-two-qubit}) has the form
\begin{equation}\label{interaction-flux}
  \hat{V} = -J_L \hat{\varphi}_A \hat{\varphi}_B\;\; {\rm with}\;\;
  J_L = \left({\hbar}/{2e}\right)^2{L_M}/\left({{L_A L_B}}\right)\,,
\end{equation}
and $E_{L, \alpha}$ is the renormalized inductive energy.

By computing $[\hat{\varphi}_\alpha, \hat{H}_\alpha^{(0)}]$, one finds
 $\omega^\alpha_{i\rightarrow f} \varphi^{\alpha}_{i\to f} = {8 E_{C,\alpha}}n^\alpha_{i\to f}$, implying that the hierarchy of $\varphi^\alpha_{i\to f}$ differs from that of $n^\alpha_{i\to f}$: $\varphi^\alpha_{0\to 1} \gtrsim \varphi^\alpha_{1\to 2}$.
Therefore, the interaction effects within the computational subspace are important. Nevertheless, this does not affect the crosstalk $\Delta_c$ since  contributions from repulsion between computational levels exactly cancel in $\Delta_c$.
 Moreover, similarly to the capacitive coupling, $\Delta_c \ll \Delta_\omega$, see Fig.~\ref{Fig-inductive}(a), allowing for the same CZ gate by driving at $\omega_{11\to 21}$. The frequency mismatch $\Delta_\omega$ now occurs primarily because of the repulsion between $|10\rangle_0$ and $|01\rangle_0$ rather than $|21\rangle_0$ and $|12\rangle_0$, see the wavy line in Fig.~\ref{Fig-fluxonium}(d). This can be seen in the matrix elements in Fig.~\ref{Fig-inductive}(b). There, $n^B_{11\to 21}$ does not grow as fast with the interaction strength as for the capacitive coupling, while $n^A_{10\to 20}$ decreases. 
The transition $|10\rangle \rightarrow |02\rangle$ acquires a relatively large matrix element, allowing for another way of activating the CZ gate. Driving this transition can be advantageous since it can be better separated in frequency from other allowed transitions [Fig.~\ref{Fig-inductive}(c)] for our choice of parameters. In this case, 
state $|10\rangle$ changes sign, and an additional 
Z$_\pi = {\rm diag}(1, e^{i\pi})$ gate applied to qubit $A$ reduces the gate to its standard form $\hat{U}_{\rm CZ}$.

We present gate errors for the inductive coupling in Figs.~\ref{Fig-inductive}(d) and \ref{Fig-inductive}(e). If the selective single-qubit drive is not possible, then driving at $\omega_{10\to 02}$ is generally a better option than driving at $\omega_{11\to 21}$. The selective drive at $\omega_{11\to 21}$  further reduces the gate error at a reasonably small interaction strength ($J_L/h < 20$ MHz) or short gate time ($t_g < 60$ ns).
The nonmonotonic behavior 
of the dashed line Fig.~\ref{Fig-inductive}(e) is explained by a decrease in separation between $\omega_{01 \rightarrow 02}$ and $\omega_{11 \rightarrow 21}$ with increasing $J_L/h$ and a relatively large value of  $n^B_{01\to 02}$.

\begin{figure}
 \includegraphics[width=0.9\columnwidth]{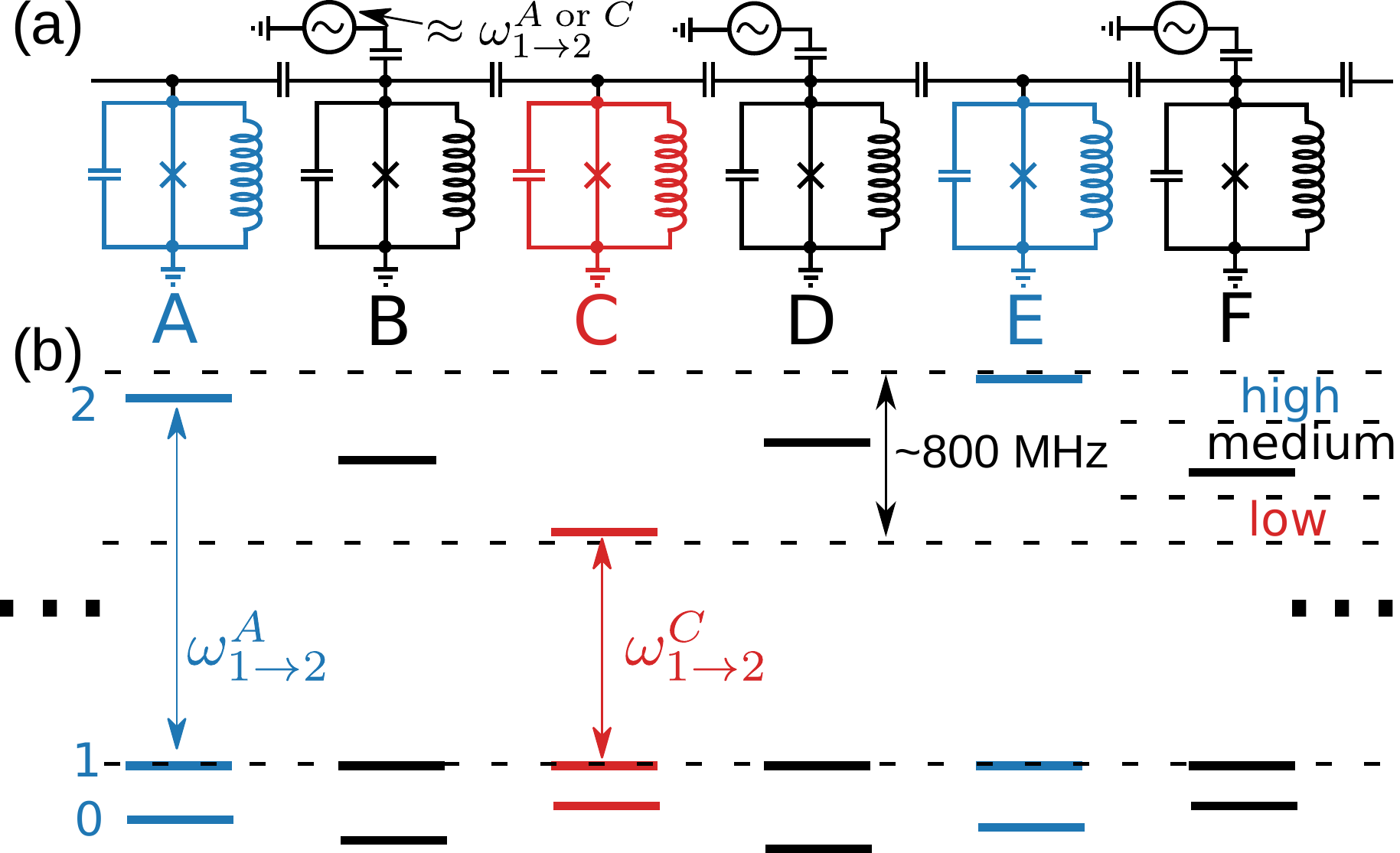}
 \caption{(Color online) (a) The circuit diagram of a chain of fluxoniums labeled as A, B, C, etc. Only those microwave   lines are shown that are used to activate CZ gates. (b) Lowest energy levels (solid lines) of the qubits shown on top with the energies  of the first excited states  aligned to emphasize the differences between the values of $\omega^\alpha_{1 \to 2}$.  }\label{Fig-scalability}
\end{figure}

\textit{Conclusion and outlook.}  
We have analyzed both inductive and capacitive interactions between two fluxoniums and found that gate fidelities of 99.9\% are possible within 100-ns gate times. The capacitive coupling scheme may be easier to realize but will require controlling the drive amplitude at both qubits, similar to the operation of the cross-resonance gate~\cite{Chow2011_prl, Sheldon2016_pra}. The inductive scheme is much less sensitive to crosstalks and state leakage but may require fine tuning of single-qubit gates or tunable couplers~\cite{Chen2014_prl}. 

Our work suggests that fluxonium qubits can be a potential upgrade to the mainstream transmons in the scalability contest. An example  is shown in Fig.~\ref{Fig-scalability}. There, qubits are divided into three groups according to  $\omega^\alpha_{1\to 2}$: high (A, E), low (C), and medium (B, D, F) transition frequencies. Every second qubit is a medium-frequency qubit, which has exactly one high- and one low-frequency neighbor. The CZ gates are realized by selectively driving medium-frequency qubits. For each of them, two corresponding resonance frequencies (e.g., $\approx\omega^A_{1\to 2}$ and $\approx\omega^C_{1 \to 2}$ for qubit $B$) are far detuned from each other. A wide parameter space allows  independent optimizations of $\omega^\alpha_{0\to 1}$ and $\omega^\alpha_{1\to 2}$; the choice of $\omega^\alpha_{0\to 1}$ is  not discussed here.

While transmon qubits face the challenges of anharmonicity-limited gate speeds and surface-material-limited lifetimes  ($T_1<100$ $\mu$s)~\cite{Wang2015_apl}, the fluxoniums at half flux quantum have millisecond energy relaxation times~\cite{Pop2014_nature} and no fundamental obstacles to achieve comparably long coherence times (the second-order flux noise  is not yet a limiting factor~\cite{Yan2016_natcomm}) together with sub-100-ns two-qubit gates. 
The extraordinary lifetime of the fluxonium $|0\rangle \to |1\rangle$ transition is partially enabled by  its very low (sub-GHz) frequency, which we believe is a virtue rather than a weakness.  The energy relaxation rate due to dielectric loss is proportionally slower at low frequencies at a constant $Q$-factor (in fact, $Q$ usually improves at lower frequencies~\cite{Braginsky1987_physlett}).   Although fluxonium qubits will operate in a relatively ``hot'' environment due to their low frequencies, a practical quantum processor in any case will employ an active qubit state initialization such as measurement feedback for either a rapid reset or suppressing nonequilibrium excitations~\cite{Riste2012_prl, Magnard2018}.  The fidelity of the state initialization directly benefits from long $T_1$.

The proposed CZ gate is made possible by the rich energy level structure of the fluxonium and separation of its well-protected memory space from strongly interacting states.  This concept is applicable to other anharmonic qubits with a hierarchy of transition matrix elements, such as variants of flux qubits~\cite{Yan2016_natcomm}.  In practice, the performance of the proposed CZ gate will depend on qubit coherence times and will likely be limited by the $T_2$ time of the $|2\rangle$ state, which is accessed during the gate.  Because the $\sim5$ GHz $|1\rangle \to |2\rangle$ transition is similar to transmon transitions, it should be possible to achieve $T_2$ time on the order of 50 $\mu$s with today's technology~\cite{Manucharyan2018}, limiting incoherent error to less than 0.1\%.  Therefore, the proposed CZ gate  provides a  promising pathway towards the long-coveted 99.9\% fidelity two-qubit gates.

\textit{Acknowledgements.}  We would like to thank Robert McDermott,  Long Nguyen, Mark Saffman, and Zhenyi Qi for fruitful discussions. We acknowledge funding from the U.S.
Army Research Office (Grants No. W911NF-15-1-0248 and No. W911NF-18-1-0146) and NSF PFC at JQI (Grant No. 1430094). Numerical simulations were performed using the
QuTiP package~\cite{Johansson2013}.

\bibliography{literature_gate}

\end{document}